# A Reo Based Solution for Engineering the Coordination Protocols for Smart Cities: A Rescue Scenario Use Case

Mohammad Reza Besharati[1]

Mohammad Izadi[2]

1  PhD Candidate, Department of Computer Engineering, Sharif University of Technology, Tehran, Iran, besharati@ce.sharif.edu

2  Associate Professor, Department of Computer Engineering, Sharif University of Technology, Tehran, Iran

## Abstract

Smart Cities, with their problems and challenges, is an emerging smart paradigm. To achieve better quality and usability levels, we need engineering solutions to support smart cities' soft-layer development. Statics, dynamics and generative semantics are involved, but segregating Coordination Protocols from the other semantics could act as a complexity management strategy to tackle the inherent complexity of smart city systems. Here we demonstrate how we could engineer the protocols layer of a smart city by using a Reo-Based solution. Using this solution, a Rescue Scenario use case is sketched and modeled.

## Introduction

Smart cities are on the way and their development, evolution, and engineering need their formal assets and languages. Each semantic space (or semantics) of involving meanings could be modeled by a separate modeling language to overcoming the complexity of the entire smart city solution.

Reo is a formal language for modeling and specification of coordination protocols [1]. An introduction of Reo syntax and behavior is mentioned in figure 1, from [2]. This language could be served as a formal modeling, verification and specification tool to engineering coordination protocols, for example see [3], [4] and [5].  Reo has a main graphical model (called Reo Circuits). Also formal semantics [6], textual notation [7], verification [8] and simulation [9] facilities are available with it and it's around ecosystem of formal coordination notations (for example see [10]).

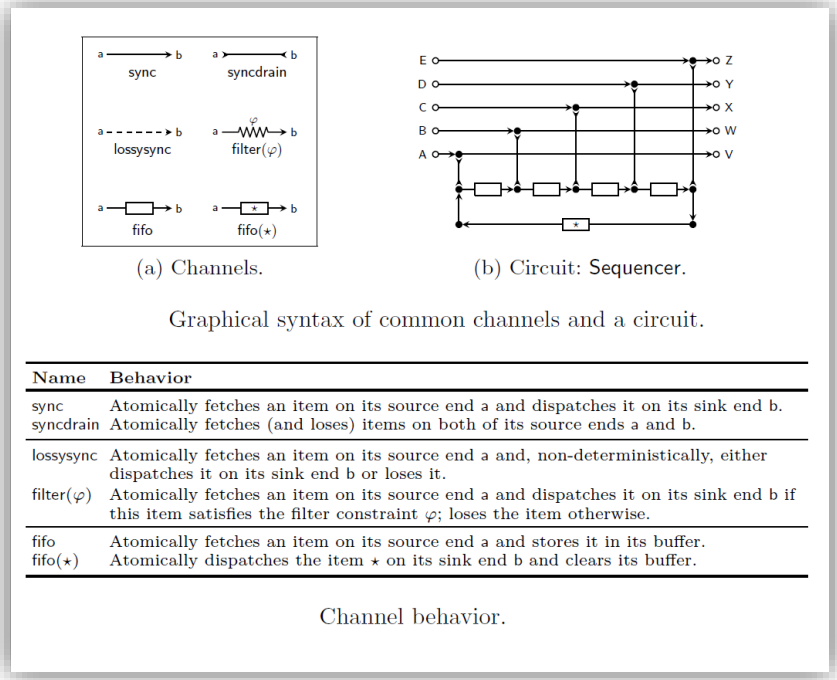

Figure 1- *About Reo Syntax and Behavior, From [2].*

## Rescue Scenario Sample

Here is a very simple rescue scenario for depicting the power of Reo language to specify, model and program the coordination protocol for rescue and emergency request management, as a control module in-part of a smart city solution:

An accident situation could be detected by an approved rescue request from citizens or sensors. The call center does the approval. Then, the incoming requests are load balanced between three Emergency-Staffs. They could activate the emergency alarm after investigating the case. Then, a Police-Staff can consider the enabling of the police alarm, and simultaneously, the Firefighting-Staff can consider the enabling of the firefighting alarm.

First, a Coordination-Flow sketch of the required protocol is provided (See Figure 2). Elements of this sketch are some common reusable components for coordination-protocols, called "connectors" [1]. Then a complete implementation (in terms of a Reo circuit which is depicting the required protocol) is derived from the previous sketch (See Figure 3).

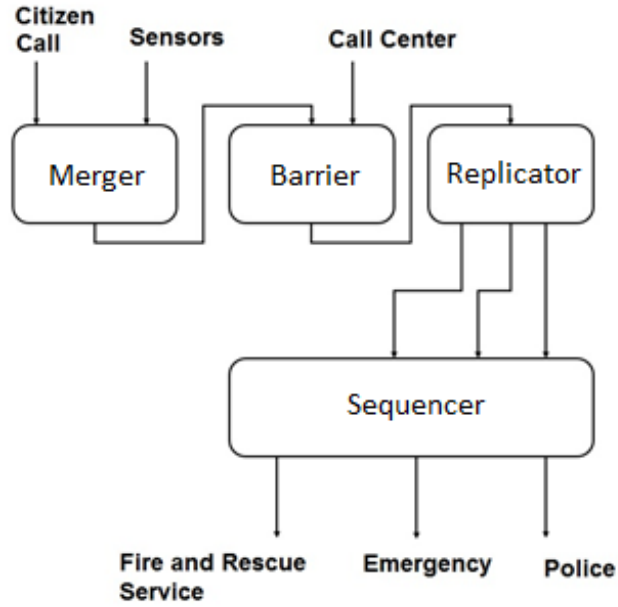

*Figure 2- Coordination-Flow Sketch, by using a mesh of some common reusable components*

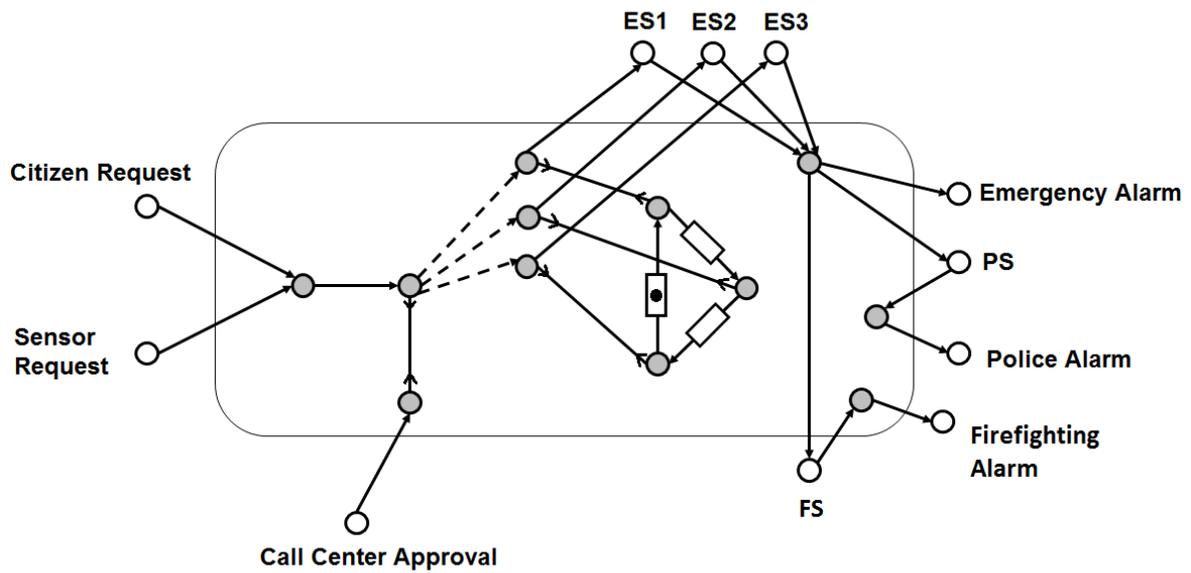

*Figure 3- A Reo Circuit implementing the intended coordination protocol for handling a recue request. ES, PS and FS mean respectively Emergency Staff, Police Staff and Firefighting Staff.*

## Semantic Logic Programming for Compliance Checking

Based on Semantic Logic [11] method and KARB solution [12], we provide a logic-based programming to implement an instance of Compliance Checking for the protocol of our example. Here is the code:

Protocol

1. AmbulanceRequest >> FireRequest >> PoliceRequest
2. PoliceRequest => P(HelicopterMisson)
3. FireRequest => P(HelicopterMission)
4. AmbulanceRequest => P(HelicopterMission)
5. HelicopterMission => BudgetConsuming

Compliance Rules

6. Forbidden((Very)BudgetConsuming)

Obligation Semantics

7. Forbidden(A) AND P(A) => Warning(P(A))
8. Forbidden(A) AND A => Failure(A)
9. Warning(P(A)) AND DoubleCheck(P(A)) => Resolved(Warning(P(A)))

Counting Semantics

10. A AND (I)A => (I+1)A
11. A =>(1)A
12. (I)A AND I>2 => P((Very)A)

Deontic Semantics

13. (A=>B) AND (P(A)) => P(B)
14. (Very)P(A) => P((Very)A)
15. P(P(A)) => P(A)

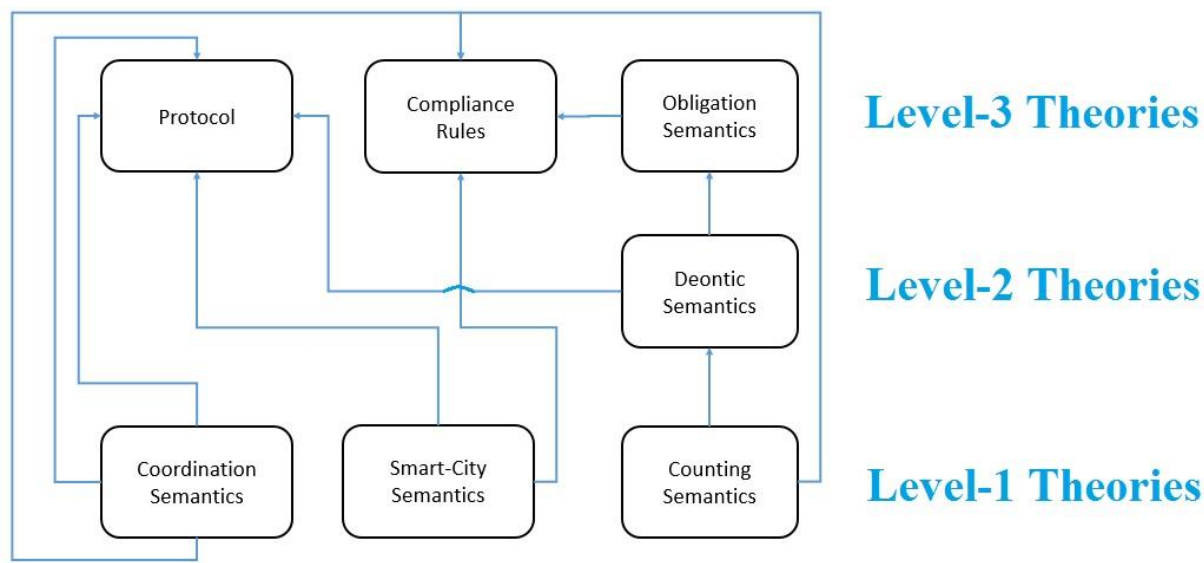

*Figure 4-* The Relation Between the involving Semantic Models.